\documentclass[prl,aps,twocolumn,showpacs,10pt]{revtex4-1}
\usepackage{amsmath,amssymb,graphicx}
\usepackage{epsfig}
\usepackage{textcomp}
\usepackage{color}
\usepackage{epstopdf}
\usepackage{array}
\usepackage[normalem]{ulem}

\newcolumntype{?}{!{\vrule width 1.5pt}}
\usepackage[final]{hyperref}
\hypersetup{pdfpagemode=UseNone,colorlinks=true,linktoc=page,pdfborder={0 0 0},linkcolor=blue,citecolor=blue}

\begin{document}
\title{Reducing the extinction risk of stochastic populations via non-demographic noise}

\author{Shay Be'er and Michael Assaf}
\email{michael.assaf@mail.huji.ac.il}

\affiliation{Racah Institute of Physics, Hebrew University
of Jerusalem, Jerusalem 91904, Israel}


\begin{abstract}

We consider non-demographic noise in the form of uncertainty in the \textit{reaction step-size}, and reveal a dramatic effect this noise may have on the stability of self-regulating populations. Employing the reaction scheme $mA\to kA$, but allowing \textit{e.g.} the product number $k$ to be a-priori unknown and sampled from a given distribution, we show that such non-demographic noise can greatly reduce the population's extinction risk compared to the fixed $k$ case. Our analysis is tested against numerical simulations, and by using empirical data of different species, we argue that certain distributions may be more evolutionary beneficial than others.
\end{abstract}

\maketitle


Dynamics of interacting stochastic populations emerge in various scientific disciplines such as physics, chemistry, population and cell biology, ecology and social sciences, and has become an important paradigm of theory of stochastic processes \cite{Bartlett,Nisbet,Horsthemke,Hanggi,Kampen,Gardiner}. A suitable framework to describe these dynamics is the so-called Markovian birth-death process, which includes a finite set of reactions of the type $mA\to kA$. Here, $A$ denotes the type of individuals participating in the reactions, and $m$ and $k$ respectively denote the number of interacting agents and products, which are a-priori known integers \cite{Gardiner}. An important subcategory of such interacting populations are self-regulating isolated populations. Even though they may dwell in a metastable state for a long time, such populations always undergo extinction with unit probability, even in the absence of environmental variations, due to an unusual chain of random events when population losses dominate over gains \cite{May,Roughgarden}. In such cases, people were mostly interested in calculating the mean time to extinction (MTE) and the probability distribution of population sizes in the long-lived metastable state prior to extinction~\cite{Andersson,Beissinger,Samoilov,PRE2010,OMII,AM17}. Notably, extinction of an isolated population may have a dramatic impact on an entire ecosystem, which is often strongly influenced by the separate abundance of various individual species, vital for the survival of the system~\cite{EcoSys1,EcoSys2,EcoSys3,EcoSys4}.

A far-less explored phenomenon in stochastic population dynamics is the effect non-demographic noise in the form of \textit{uncertainty} in the reaction \textit{step-size} may have on the dynamics of interest, and in particular, on rare-event statistics. An example for such reaction step-size noise (RSSN) is the process $A\to kA$, where the reaction product number $k=1,2,3,\dots$ (\textit{e.g}, the different litter sizes) is an a-priori \textit{unknown} integer, and is sampled from a (given) step-size distribution (SSD), $D(k)$. This means that in a stochastic realization of this process, at each time step, once a birth event is chosen, the jump size has to be also chosen from a given SSD~\cite{Langevin}. Also, we refer to RSSN as non-demographic noise, since the distribution of product numbers $D(k)$ depends on environmental factors that are extrinsic to the population of interest.

This type of uncertainty, or noise, appears in a wide variety of scientific areas including physics of nuclear reactors~\cite{Stacey}, population biology and ecology~\cite{OMII,Goel}, viral dynamics~\cite{Krapivsky,Sinitstyn}, and cell biology~\cite{Paulsson,SwainI,Swain,1945}.  However, apart from particular examples, see \textit{e.g.}, \cite{Paulsson,Krapivsky,Sinitstyn}, RSSN has not been rigorously studied. Recently, we have considered the effect of RSSN when incorporated in the influx and reproduction processes~\cite{OURS-BI,OURS-JSTAT}. However, the fundamental question of how adding generic noise in the reaction step size  influences the statistics of interest relative to the fixed $k$ case, has not been addressed so far.

\begin{figure}[b]
\includegraphics[trim = .3in 2.1in 1in .15in,clip,width=3.4in]{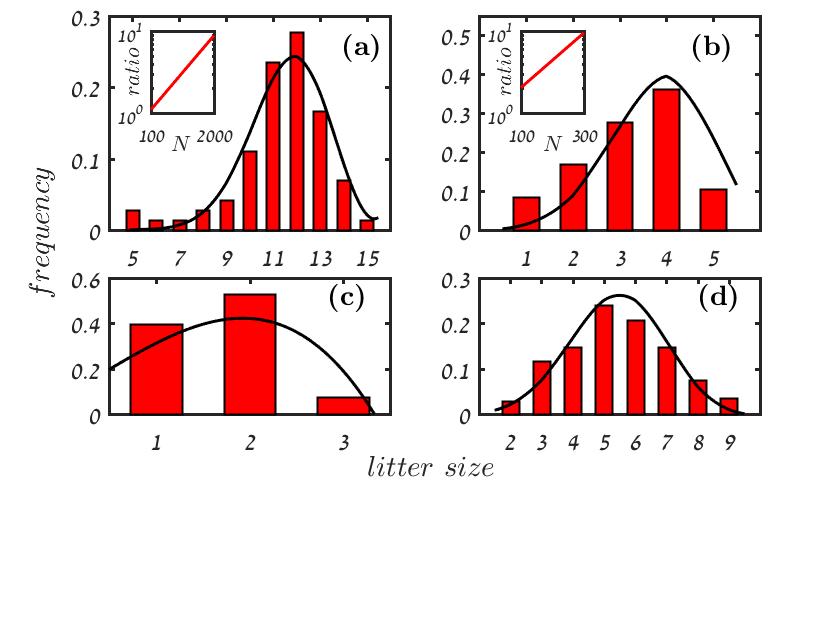}
\caption{(Color online) frequency of appearance of different litter sizes. Bars correspond to empirical data for laboratory mouse~\cite{mice} (a), guinea pig~\cite{guinea} (b), spotted hyena~\cite{heynas} (c), and wild boar~\cite{boar} (d). Lines correspond to binomial distribution fits (see text) with best fit parameters of $m=15$ and $p=0.775$ (a), $m=5$ and $p=0.75$ (b), $m=3$ and $p=0.575$ (c), and $m=9$ and $p=0.6$ (d). Insets correspond to the ratio $\tau^\mathrm{emp}/\tau^\mathrm{ref}$ as a function of $N$ with $B=2$, see text.}
\label{fig-a}
\end{figure}

In this paper we reveal a dramatic impact RSSN can have on the stability and persistence of the population. Using the WKB formalism, both in real and momentum space, we show that depending on the SSD's characteristics, uncertainty in the step size can either increase or decrease the population's extinction risk and MTE, compared to the case of a fixed product number. In particular, close to bifurcation we analytically find the threshold which depends on the SSD's first and second moments, beyond which RSSN increases the stability of the long-lived metastable state and thereby, decreases the population's extinction risk. Moreover, we demonstrate how this non-demographic noise can be tuned to achieve maximal reduction in the extinction risk. Finally, we propose that this phenomenon of reducing the extinction risk via noise may indicate that RSSN is a natural mechanism adopted by real species in order to enhance their fitness and persistence, see Fig. \ref{fig-a}. Thus, it is evolutionary beneficial to the entire ecosystem depending on these species.


We begin with a simple model which is a bursty variant of the well-known logistic model, defined by the reactions
\begin{equation}
	\label{a}
	A\xrightarrow{\lambda_n}A+kA \ \ (k=0,1,2,\dots), \ \ \ \ A\xrightarrow{\mu_n}\emptyset.
\end{equation}
Here, $\lambda_n=Bn(1-n/N)[D(k)/\langle k\rangle]$ is the birth rate~\cite{RE}, $\mu_n=n$ is the death rate, $n$ is the size of the population, $B\gtrsim1$ is the per-capita average reproduction rate, $N\gg1$ is the typical population size in the long-lived metastable state prior to extinction (see below), and time is rescaled by the death rate. Also, $k$ is the offspring number per birth event, a-priori unknown and drawn from an arbitrary, normalized SSD, $D(k)$, with the mean value, $\langle k\rangle$, and standard deviation, $\sigma_k$.

The deterministic rate equation describing the time-evolution of the mean population size reads $\dot{\bar{n}}=\lambda_{\bar{n}}-\mu_{\bar{n}}=\bar{n}\left[B(1-\bar{n}/N)-1\right]$. The steady-state solutions are the repelling fixed point, $n=0$, and the attracting fixed point, $n_s=N(1-1/B)\gg1$, indicating that the population size flows into $n_s$ starting from any nonzero population size. Yet, owing to the existence of an absorbing state at $n=0$, demographic fluctuations ultimately cause the population to go extinct with a unit probability. In order to incorporate demographic noise and to study the extinction statistics, we write down the master equation describing the time history of $P_n(t)$ - the probability to find $n$ individuals at time $t$:
\begin{align}
	\label{c}
	\dot{P}_n = & \frac{B}{N\langle k\rangle} \left[\sum_{k=0}^{n-1}(n-k)(N-n+k)D(k)P_{n-k}\right. \\ & \left.-\sum_{k=0}^{\infty}n(N-n)D(k)P_n\right]+(n+1)P_{n+1}-nP_n. \nonumber
\end{align}
Note that in the first term on the right hand side of Eq. (\ref{c}), the summation can be formally extended up to infinity since it is assumed that $P_{n<0}=0$. The summation in the second term yields $-n(N-n)P_n$.

At times much larger than the system's deterministic relaxation time, $t_r$, once the system has already settled in the long-lived metastable state, the probability distribution around the metastable state slowly decays in time  $P_{n>0}(t\gg t_r)\simeq\pi_ne^{-t/\tau}$, while the extinction probability grows, $P_0(t\gg t_r)\simeq 1-e^{-t/\tau}$. Here, $\tau$ denotes the MTE, and $\pi_n$ denotes the quasi-stationary distribution which is the shape of the metastable state~\cite{PRE2010,AM17}. In order to calculate the MTE and quasi-stationary distribution in the leading order, see SM, we employ the real-space WKB ansatz, $\pi_n\equiv\pi(q)\simeq e^{-N\mathcal{S}(q)}$~\cite{Dykman,KS,explosion,EK,SBDL,PRE2010,AM17}, where $\mathcal{S}(q)$  is called the action function and $q=n/N$ is the rescaled coordinate. Finding the subleading-order terms, however, requires using the probability generating function formalism~\cite{Gardiner} in conjunction with the WKB approach~\cite{AM17}. Doing so, see details in the SM, the MTE, $\tau$, is found to be
\begin{equation}	
	\label{f}
	\tau=\frac{B}{(B-1)(e^{-\mathcal{S}'(0)}-1)} \ \sqrt{\frac{2\pi|\mathcal{S}''(0)|}{N}}e^{N\mathcal{S}(0)}.
\end{equation}
Note that close to the bifurcation, $B-1\ll 1$, this result considerably simplifies. Here, the attracting fixed point, $q_s=n_s/N=1-1/B$, becomes close to the extinction point, $q=0$, such that $q_s\ll 1$. In this case~(\ref{f}) becomes
\begin{equation}
	\label{g}
	\tau=\frac{\sqrt{\pi(1+\gamma)}}{\sqrt{N}(B-1)^2}e^{\frac{N(B-1)^2}{1+\gamma}}, \ \ \ \gamma\equiv\langle k \rangle+\frac{\sigma_k^2}{\langle k\rangle},
\end{equation}
see SM for details. Since it includes subleading-order corrections, this result is applicable as long as the strong inequality $N^{-1/2}\ll B-1\ll N^{-1/3}$ holds.

\begin{figure}[b]
\includegraphics[trim = .5in 3.85in 2.6in 3.5in,clip,width=3.4in]{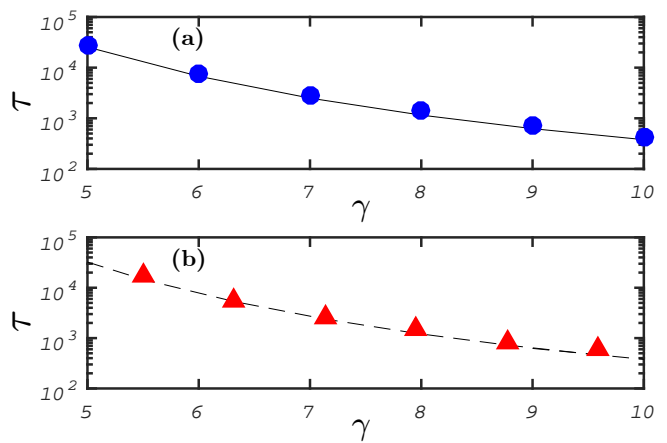}
\caption{(Color online) MTE as a function of $\gamma$. Lines are obtained by numerically evaluating Eq.~(\ref{f}). Markers correspond to numerical Monte Carlo simulation results. Panel (a) and (b) correspond to the $K$-step and binomial cases, respectively. Parameters: $B=2$, $N=150$, and $m=10$.}
\label{fig-b}
\end{figure}

In addition to the MTE, the quasi-stationary distribution can also be found once the SSD is explicitly specified, see SM. An important quantity that represents the overall noise magnitude (both demographic and non-demographic) and strongly affects the MTE is the observed variance of the quasi-stationary distribution, $\sigma_\mathrm{obs}$ \cite{OURS-BI,OURS-JSTAT}, which can be found for any SSD
\begin{equation}
	\label{e}
	\sigma_\mathrm{obs}^2=N\frac{(1+\gamma)}{2B},
\end{equation}
where $\gamma$ is defined in Eq.~(\ref{g}), and depends on the first two moments of the SSD. As one can see, both $\sigma_\mathrm{obs}$ and $\tau$ in the bifurcation limit, are governed by the factor $\gamma$.

In order to quantify the effect of RSSN on the extinction dynamics, we compare these results obtained in the presence of RSSN, with a simple $K$-step reference model where $D^\mathrm{ref}(k) = \delta_{k,K}$. Here, the step size is fixed and equals $K$, corresponding to, \textit{e.g.}, the most frequent litter size. Naturally, however, the litter size is known to be distributed around some value $K$. Thus, a more suitable choice would be some SSD with a finite width, see Fig.~\ref{fig-a}.
As stated, each SSD is associated with a different $\gamma$ factor (\ref{g}), where close to bifurcation, the relation between the latter and $\gamma^\mathrm{ref}=K$ defines whether RSSN increases or decreases the population's extinction risk~\cite{nonbifur}. Indeed, for $\gamma<\gamma^\mathrm{ref}$ we have $\sigma_\mathrm{obs}<\sigma_\mathrm{obs}^\mathrm{ref}$ and $\tau>\tau^\mathrm{ref}$ and vice versa. Importantly, this effect of decreasing the extinction risk of the population via RSSN can be enhanced by choosing optimal parameters for the SSD, see below.

For concreteness we proceed by choosing a binomial distribution for the SSD: $D^\mathrm{bin}(k) = {m \choose k}p^k(1-p)^{m-k}$, where ${m \choose k}=m!/[k!(m-k)!]$ represent the number of ways to choose $k$ out of $m$. Here, the burst size, $k$, which satisfies $0\leq k\leq m$, is an integer representing the number of successes -- the actual number of offspring produced -- out of a maximum possible number of $m$ offspring. To compare the bursty and reference models, we tune the success probability of a single event, $0<p<1$, in such a way that the \textit{mode}, \textit{i.e.} the most probable value of the SSD, is equal to $K$. This can be achieved by choosing $p=(K+1/2)/(m+1)$. Figure \ref{fig-a} presents data deduced from field observations on four different animal species which empirically determined the frequency of appearance of different litter sizes. Fitting the empirical data with the binomial distribution shows that the latter is an adequate candidate to describe the litter size distribution.

In Fig.~\ref{fig-b} we compare theoretical and simulation results for the MTE as a function of $\gamma$, see Eq.~(\ref{g}), for the reference and binomial SSDs. The theoretical results, found by numerically evaluating Eq.~(\ref{f}), agree well with results of Monte-Carlo simulations~\cite{Gillespie}.

\begin{figure}[t]
\includegraphics[trim = .95in 5.4in 1.6in 2.9in,clip,width=3.4in]{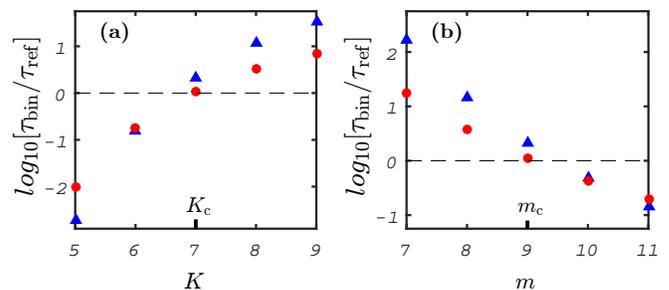}
\caption{(Color online) Logarithm of the ratio between the binomial and reference MTEs versus $m$ and $K$. ({\color{blue}$\blacktriangle$}) markers are obtained by numerically evaluating Eq.~(\ref{f}) with $B=2$ and $N=2000$. ({\color{red}$\bullet$}) markers correspond to Eq.(\ref{g}) with $B=1.15$ and $N=25000$. Panel (a): $m=9$ (for which $K_\mathrm{c}=7$). Panel (b): $K=7$ (for which $m_\mathrm{c}=9$).}
\label{fig-c}
\end{figure}

We now use the binomial SSD in order to find the critical value of $m$ below which RSSN decreases the extinction risk compared to the reference SSD. To this end we solve the equation $\gamma^\mathrm{bin}=\gamma^\mathrm{ref}$, with $\gamma^\mathrm{bin}=1+(K+1/2)(m-1)/(m+1)$ [see Eq.~(\ref{g})]. This yields the critical values $m=m_\mathrm{c}(K)=(4K-1)/3$ and $K=K_\mathrm{c}(m)=(3m+1)/4$. As a result, for $m<m_\mathrm{c}(K)$, we have $\sigma_\mathrm{obs}^\mathrm{bin}<\sigma_\mathrm{obs}^\mathrm{ref}$, and vice versa. Close to the bifurcation limit this indicates that for $m<m_\mathrm{c}(K)$ [or $K>K_\mathrm{c}(m)$], RSSN decreases the extinction risk, namely $\tau^\mathrm{bin}>\tau^\mathrm{ref}$~\cite{nonbifur}. In Fig. \ref{fig-c} we numerically confirm that this onset also holds well beyond the bifurcation limit. Here, using our theoretical results [Eqs.~(\ref{f}) and (\ref{g})], we compared between the binomial and $K$-step SSDs close to bifurcation and also in the non-bifurcation regime. Our results confirm that for $K>K_\mathrm{c}(m)=(3m+1)/4$ (left panel) or $m<m_\mathrm{c}(K)=(4k-1)/3$ (right panel), RSSN increases the population's stability compared to the $K$-step reference model. One also observes that this behavior also holds well into the non-bifurcation regime.

An intuitive way to understand this value of $m=m_\mathrm{c}$ (or $K=K_\mathrm{c}$) is to look at the SSD's \textit{skewness}, $\hat{\mu}_3$, which is the third standardized moment of the SSD. For $p=(K+1/2)/(m+1)$ and $m\leq m_\mathrm{c}$, the skewness of the binomial SSD is always negative. A negatively-skewed distribution typically has a left tail which is fatter than the right tail.  We have established by now that in order for the RSSN to decrease the extinction risk compared to the $K$-step reference case, we must have $\gamma^\mathrm{bin}<\gamma^\mathrm{ref}$. To achieve this, a negatively-skewed SSD, for which the mean is smaller than the mode, is required. Furthermore, the SSD has to be sufficiently (negatively) skewed, since the variance of the SSD also contributes to the increase of the parameter $\gamma$. Notably, since the average birth rate of the negatively-skewed distribution is larger than that of the reference model, such RSSN increases the overall frequency of  birth events. Therefore, such RSSN decreases the probability for a chain of death events leading to extinction, compared with the $K$-step reference model, and thus, decreases the extinction risk of the population.

\begin{figure}[t]
\includegraphics[trim = .6in 3in .5in 2.7in,clip,width=3.4in]{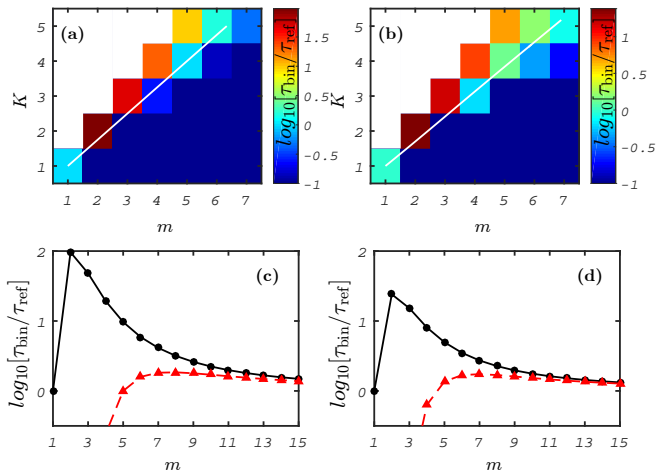}
\caption{(Color online) 3D plots of the logarithm of the ratio between the binomial and reference MTEs as a function of $m$ and $K$ in the bifurcation (a) and non-bifurcation (b) regimes. In (a) and (b) the results are respectively obtained using Eq.~(\ref{g}), and by numerically evaluating Eq.~(\ref{f}). Along the solid lines in panels (a) and (b), $\tau^\mathrm{bin}=\tau^\mathrm{ref}$. In panels (c) and (d) we plot sections of panels (a) and (b), putting $K=m-R$, for $R=0,1$. The maxima in panel (a), obtained at $m_\mathrm{opt}=2,8$, are accurately predicted by Eq.~(\ref{k}). In panel (b) the maxima, at $m_\mathrm{opt}=2,7$, are still closely predicted by Eq.~(\ref{k}). Here $B=1.125$ and $N=15000$, for panels (a) and (c), and $B=2$ and $N=400$, for panels (b) and (d).}
\label{fig-d}
\end{figure}

Having found the onset for the stability enhancement of the population due to RSSN, we can also find the optimal SSD parameters which \textit{maximize} the ratio $\tau^\mathrm{bin}/\tau^\mathrm{ref}$, namely maximize the effect due to RSSN. Let us compute this ratio analytically close to bifurcation, where in the following we will also refer to the non-bifurcation regime. Since $m\geq K$, it is convenient to analyze the ratio $\tau^\mathrm{bin}/\tau^\mathrm{ref}$ as a function of $m$ on curves $K=m-R$ with $R=0,1,2,\dots$, which span the entire space of $(m,K)$. Using Eq.~(\ref{g}), differentiating the ratio with respect to $m$, and expanding the result in $m\gg 1$~\cite{largem}, we find the optimal $m$ which maximizes the ratio
\begin{equation}
	\label{k}
	m_\mathrm{opt}\simeq \displaystyle \lfloor 2.04+5.64R \rceil, \ \ \ K_\mathrm{opt}=m_\mathrm{opt}-R,
\end{equation}
where, $\displaystyle \lfloor \rceil$ denotes the nearest integer. This result indicates that for a given $R=m-K$, RSSN has the strongest effect for SSDs with a distinct relation between $m$ and $K$:  $m=m_\mathrm{opt}(R)$ [or $K=K_\mathrm{opt}(R)$]. Figures~\ref{fig-d}{\color{blue}a} and \ref{fig-d}{\color{blue}b} show the ratio $\tau^\mathrm{bin}/\tau^\mathrm{ref}$ as a function of $m$ and $K$ in the bifurcation and non-bifurcation regimes, respectively, where in both panels a very similar qualitative behavior is observed. The solid lines represent the case for which the ratio equals $1$. In Figs.~\ref{fig-d}{\color{blue}c,d} we show equal-$R$ sections of Figs.~\ref{fig-d}{\color{blue}a,b}, respectively. One can see that for a given $R$ value, each curve receives a maximum for a given value of $m$, which, close to bifurcation, is accurately predicted by Eq.~(\ref{k}). Moreover, while Eq.~(\ref{k}) formally applies only in the bifurcation limit, its predictions also hold well within the non-bifurcation regime (panel d). Finally, note that because of the proximity of the mode and mean in the case of the binomial distribution, the ratio $\tau^\mathrm{bin}/\tau^\mathrm{ref}$ is always below $1$ for $R\geq 2$. To overcome this limitation, one can use instead a generalized version of the binomial SSD in the form of the beta-binomial SSD~\cite{betabin,Vilk}.


The results we have obtained so far are generic and model-independent. To illustrate this point, we briefly present a different model of birth-death-competition, which includes the following set of reactions
\begin{equation}
	\label{l}
		A\xrightarrow{\lambda_n}2A, \ \ A\xrightarrow{\mu_n^{(1)}}\emptyset, \ \ M A\xrightarrow{\mu_n^{(2)}}(M-k)A.
\end{equation}
Here, $\lambda_n=B n$ is the birth rate, $\mu_n^{(1)}=n$ is the linear death rate, and $\mu_n^{(2)}=(M!/N^{M-1}){n \choose M}[D(k)/\langle k\rangle]$ is the competition rate. It corresponds to a scenario in which a competition between $M$ individuals results in the removal or death of $k=0,\dots,M$ of them with a probability of $D(k)$. All other parameters are similar to those of the reproduction model. Note, that the rate of the non-linear competition reaction was chosen in such a way that the corresponding rate equation of this model coincides with that of the reference model with $D^\mathrm{ref}(k) = \delta_{k,K}$, see SM.

As in the previous model, here an initial population first relaxes into the long-lived metastable state, from which it ultimately goes extinct via a large fluctuation after an exponentially long waiting time. The calculations of the quasi-stationary distribution and MTE can be done along the same lines of the bursty reproduction model and are presented in the SM. Importantly, as shown in the SM, in this model one also finds that in some region of parameters, RSSN can decrease the population's extinction risk compared to the reference model, and it is possible to tune the SSD's parameters in order to maximize the population's persistence due to RSSN.


In this paper we have demonstrated a dramatic effect reaction step-size noise (RSSN) may have on the stability of an isolated self-regulating population. We have found that in a certain regime of parameters associated with the step-size distribution (SSD), RSSN can decrease the population's extinction risk compared to a model with a fixed burst size. These results indicate that remarkably, non-demographic noise in the form of uncertainty in the reaction product number, can increase the persistence of a stochastic population. We have also shown how this effect can be maximized by tuning the SSD's parameters. Our calculations were carried out on two different models of population dynamics, and while we have used the binomial distribution as a prototypical example for the SSD, we have checked that these effects described above also hold for other SSDs such as the triangular or negative-binomial distributions.

As stated, the position of the SSD's mean with respect to its mode is key in understanding this effect. As the mean becomes lower than the mode, the overall frequency of birth events increases compared to that of the reference $K$-step model. Therefore, the probability for a series of successive death events that drives the population to extinction is \textit{decreased} when the SSD's mean is sufficiently low, compared to the reference model. Thus, by taking a sufficiently negatively-skewed SSD, $\gamma$ (see text) can be reduced below the threshold of the reference model, and thereby increase the population's stability.

Finally, our analysis allows us to explore the effect of stability enhancement in realistic scenarios, by using real-life empirical data of litter-size distributions of four different animal species presented in Fig.~\ref{fig-a}. To calculate the expected MTE, we arbitrarily choose $B=2$ and numerically evaluate Eq.~(\ref{f}) to obtain $\tau^\mathrm{emp}$. This is done by using the empirical litter size distributions shown in Fig.~\ref{fig-a}. As before, the reference MTE, $\tau^\mathrm{ref}$, corresponds to a scenario of constant number of offspring which equals the distribution's mode. Performing the calculations, we find $\ln[\tau^\mathrm{emp}/\tau^\mathrm{ref}]\simeq0.0011 N$ for laboratory mice, $0.0077 N$ for  guinea pigs, $0.0061 N$ for spotted hyena, and $-0.0078 N$ for wild boars, see insets in Fig.~\ref{fig-a}.  Thus, \textit{i.e.}, choosing $N=500$, the lifetimes of laboratory mice, guinea pigs and spotted hyenas grow by a factor of $1.7$, $47$, and $21.1$, respectively, due to RSSN, whereas, only for the litter-size distribution of wild boar we find a decline by a factor of $0.02$ in the MTE for this choice of parameters. As demonstrated by our theory, we find that by having such negatively-skewed litter-size distribution (with a $\gamma$ factor smaller than that of the reference model), these species are able to increase their stability and lifetime compared to that of the reference model, whereas positive skewness has the opposite effect.

As a result, it is likely that species capable of adopting such behavior are more likely to decrease their extinction risk and to have a higher fitness. However, further study of this phenomenon is required, especially in ``real-life" biological and ecological systems.

\vspace{8.10mm}
\large
\begin{center}
\textbf{Appendix}
\end{center}

\begin{center}
	{\textbf{A. Bursty reproduction model -- mean time to extinction calculation}}
\end{center}

\vspace{1mm}

In this section we provide detailed calculations of the leading and subleading-order contributions to the mean time to extinction (MTE) in the bursty reproduction model presented in the main text.


\vspace{4mm}

\begin{center}
	{\textit{1. Leading order calculations}}
\end{center}

\vspace{2mm}

Here we derive the leading-order MTE employing the ``real-space" WKB theory, see \textit{e.g.}, Refs.~\cite{Dykman,KS,explosion,EK,PRE2010}. Our starting point is the master equation [Eq.~(2) in the main text]. To this end we use the metastable ansatz, $P_{n>0}(t\gg t_r)\simeq\pi_ne^{-t/\tau}$, and $P_0(t\gg t_r)\simeq 1-e^{-t/\tau}$, which is valid at times much larger than the typical relaxation time of the system. Here $\tau$ denotes the MTE, and $\pi_n$ denotes the quasi-stationary distribution of population sizes. Defining the rescaled coordinate $q=n/N$, where $N\gg 1$ is the typical system size in the metastable state, and employing the WKB ansatz for the quasi-stationary distribution, $\pi_n\equiv\pi(q)\simeq e^{-N\mathcal{S}(q)}$~\cite{Dykman,KS,explosion,EK,PRE2010}, where $\mathcal{S}(q)$  is the action function, we arrive at a stationary Hamilton-Jacobi equation $\mathcal{H}(q,p)=0$, with the Hamiltonian
\begin{equation}
	\label{A.1.a}
	\mathcal{H}(q,p)=(e^p-1)B q\left[(1-q)f(p)-1/(Be^p)\right].
\end{equation}
Here, we have introduced the momentum $p=\mathcal{S}'(q)$, and defined $f(p)=[\sum_{k=0}^\infty e^{kp}D(k)-1]/[\langle k\rangle (e^p-1)]$~\cite{OURS-BI,OURS-JSTAT}. The latter is related to the moment generating function of the step-size distribution (SSD).

\begin{figure}[h!]
\centering
\includegraphics[trim = .5in 2.85in 2.6in 6in,clip,width=3.0in]{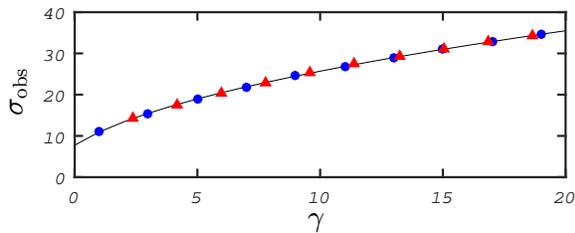}
\caption{(Color online) Observed standard deviation of the quasi-stationary distribution versus $\gamma$. The solid line corresponds to the theoretical result [Eq.~(\ref{A.1.d})]. The ({\color{red}$\blacktriangle$}) and ({\color{blue}$\bullet$}) markers correspond to numerical Monte Carlo simulation results corresponding to the binomial and reference cases (defined in the main text), respectively. In order to determine different $\gamma$ values, we kept the parameter $m$ constant and changed the parameter $K$. Parameters: $B=2.5$, $N=300$, and $m=20$.}
\label{fig-sa}
\end{figure}

The non-trivial zero-energy trajectory of the Hamiltonian (\ref{A.1.a}), called the activation trajectory, encodes the statistics of rare events~\cite{Dykman}. It is given by
\begin{equation}
	\label{A.1.b}
	q_a(p)=1-\frac{1}{Be^pf(p)},
\end{equation}
and allows for the calculation of the action according to
\begin{equation}
	\label{A.1.c}
	\mathcal{S}(q) =\int_{q_s}^qp_a(q')dq'=\int_0^{p_a(q)}p'\frac{dq_a(p')}{dp'}dp',
\end{equation}
where we have removed the arbitrary constant by defining $\mathcal{S}(q_s)=0$. Having found the action we now have the leading-order quasi-stationary distribution up to a normalization constant, which can be found by using a Gaussian approximation in the vicinity of $q_s$, see \textit{e.g.}, Refs.~\cite{EK,PRE2010}. While finding the quasi-stationary distribution requires the explicit knowledge of the SSD, the observed variance of the distribution, $\sigma_\mathrm{obs}^2=N/\mathcal{S}''(q_s)$, can be found for any SSD and is given by
\begin{equation}
	\label{A.1.d}
	\sigma_\mathrm{obs}^2=N\frac{(1+\gamma)}{2B}, \ \ \ \gamma\equiv\langle k \rangle+\frac{\sigma_k^2}{\langle k\rangle}.
\end{equation}
Here, we have used the relation $dp_a(q)/dq|_{q=q_s}=1/[dq_a(p)/dp|_{p=0}]$ and employed $L'H\hat{o}pital$'s rule to evaluate $f(0)$ and $f'(0)$. Figure \ref{fig-sa} shows the theoretical result for $\sigma_\mathrm{obs}$ as a function of $\gamma$, given by Eq.~(\ref{A.1.d}). Here we compare, for the binomial and reference SSDs (see main text), the analytical result with numerical Monte Carlo simulation results using the Gillespie algorithm~\cite{Gillespie}, and excellent agreement is observed.

Once the quasi-stationary distribution has been found, the MTE is given in the leading order the by $\tau\sim \pi_1^{-1}$~\cite{PRE2010}, which yields
\begin{equation}
	\label{A.1.e}
	\tau\simeq e^{N\mathcal{S}(0)}.
\end{equation}
Note that evaluating the action at the extinction point requires evaluating the ``fluctuational momentum", defined as $p_f=p_a(0)$, which can be done by solving the equation $q_a(p_f)=0$. This non-zero momentum which lies on the activation trajectory differs from the momentum at the mean-field extinction point (or at any deterministic fixed point) which, by definition, has to be zero.

Finally, we note that the MTE can also be calculated by going through a different route: by calculating the mean extinction rate for a given step size $k$, and averaging the result over $k$ with the corresponding weights $D(k)$. This averaging can be done by using \textit{e.g.} the saddle-point approximation. However, in the method presented above the obtained Hamiltonian [Eq.~(\ref{A.1.a})] already represents the exact weighted average over all possible step sizes, by introducing the function $f(p)$. Therefore, our method does not require the final averaging step over the extinction rates, and thus, is simpler.

\vspace{6mm}


\begin{center}
	{\textit{2. Sub-leading order calculations}}
\end{center}

\vspace{2mm}

Here we outline the main steps of the calculation of the sub-leading order correction to the MTE. Our calculation closely follows the Appendix of Ref.~\cite{OURS-JSTAT} where a similar model has been analyzed. Here, unlike the leading-order calculation, we use the so-called ``momentum-space approach", see \textit{e.g.} Refs.~\cite{Elgart,MS} which employs the generating function technique in conjunction with the spectral formalism~\cite{MS,spectral_I,spectral_II,spectral_III}. The probability generating function is defined as~\cite{Gardiner}
\begin{equation}
	\label{A.2.a}
	G=\sum_{n=0}^\infty \mathfrak{p}^n P_n,
\end{equation}
where $\mathfrak{p}$ is the conjugate momentum to the coordinate $\mathfrak{q}$ in the momentum space~\cite{Elgart,MS,Footnote1}. Note that the momentum-space $(\mathfrak{q},\mathfrak{p})$ is related to the real-space $(q,p)$ by a canonical transformation~\cite{Elgart,MS}. Multiplying the master equation [Eq.~(2) in the main text] by $p^n$ and summing over all $n$, we arrive at an evolution equation for the probability generating function \cite{Elgart,MS} which reads
\begin{eqnarray}
	\label{A.2.b}
	\partial_tG&=&(\mathfrak{p}-1)\left\{\left[B \mathfrak{p}\tilde{f}(\mathfrak{p})\left(1-\frac{1}{N}\right)-1\right]\partial_\mathfrak{p}G \right.\nonumber\\
&-&\left.\frac{B}{N}\mathfrak{p}^2 \tilde{f}(\mathfrak{p})\partial_{\mathfrak{p}\mathfrak{p}}G\right\},
\end{eqnarray}
where we have defined $\tilde{f}(\mathfrak{p})=[\sum_{k=0}^{\infty}\mathfrak{p}^kD(k)-1]/[\langle k\rangle(\mathfrak{p}-1)]$.

We now expand the probability generating function (\ref{A.2.a}) in the yet unknown eigenvalues and eigenmodes. Focusing on times much larger than the typical relaxation time when the higher eigenmodes have already decayed, we can write~\cite{MS,spectral_I,spectral_II,spectral_III}
\begin{equation}
	\label{A.2.c}
	G(\mathfrak{p},t)\simeq1-\varphi(\mathfrak{p})e^{-Et}.
\end{equation}
Here, the first term on the right hand side corresponds to extinction at $t\to\infty$, $\varphi(\mathfrak{p})$ is the lowest-excited eigenmode, and $E$ is the lowest-excited eigenvalue, which equals the inverse of the MTE~\cite{MS,spectral_I,spectral_II,spectral_III}. Substituting ansatz~(\ref{A.2.c}) into the evolution equation~(\ref{A.2.b}), one obtains a second order ordinary differential equation for $\varphi(\mathfrak{p})$
\begin{eqnarray}
	\label{A.2.d}
&&	\frac{B}{N}\mathfrak{p}^2(\mathfrak{p}-1)\tilde{f}(\mathfrak{p})\varphi''(\mathfrak{p}) +(\mathfrak{p}-1)\left[1-B \mathfrak{p}\tilde{f}(\mathfrak{p})\right.\nonumber\\
&&\left.\times\left(1-\frac{1}{N}\right)\right]\varphi'(\mathfrak{p}) -E\varphi(\mathfrak{p}) = 0,
\end{eqnarray}
where prime denotes differentiation with respect to $\mathfrak{p}$. This equation cannot be solved exactly and therefore, we solve it in two different regions and match the solutions to find the MTE. Following the Appendix in Ref.~\cite{OURS-JSTAT}, in the first region, $0\leq\mathfrak{p}<1$ (but not too close to $\mathfrak{p}=1$), where the solution is slowly varying, we perturbatively solve the equation using the ansatz $\varphi(\mathfrak{p})=1+\delta\varphi(\mathfrak{p})$, where $\delta\varphi(\mathfrak{p})\ll 1$. In the second, boundary-layer region, $N^{-1}\ll 1-\mathfrak{p}\ll 1$, we find the rapidly-varying solution by neglecting the exponentially small term $E\varphi(\mathfrak{p})$ in Eq.~(\ref{A.2.d}). By doing so, see Appendix of Ref.~\cite{OURS-JSTAT} for details, and matching the solutions in their joint region of applicability, we find the MTE, which yields~\cite{canonical}
\begin{equation}	
	\label{A.2.e}
	\tau=\frac{B}{(B-1)(e^{-\mathcal{S}^{'}(0)}-1)} \ \sqrt{\frac{2\pi|\mathcal{S}^{''}(0)|}{N}}e^{N\mathcal{S}(0)}.
\end{equation}

\begin{figure}[h!]
\centering
\includegraphics[trim = 0.65in 5.3in 1.25in 2.65in,clip,width=3.4in]{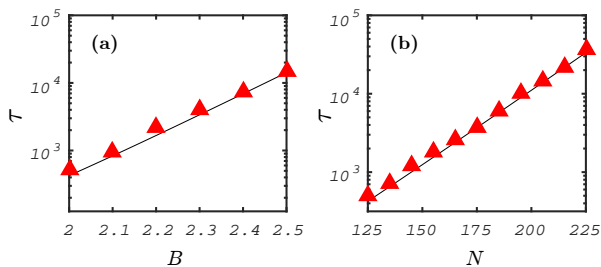}
\caption{(Color online) MTE for the bursty-reproduction model, using a binomial SSD (defined in the main text), as a function of the model parameters. Lines correspond to the theoretical MTE (\ref{A.2.e}); Markers correspond to numerical results using Monte Carlo simulations. Panel (a): $N=125$; Panel (b): $B=2$; Both panels: $K=8$ and $m=10$.}
\label{fig-sb}
\end{figure}

Figure \ref{fig-sb} presents the MTE for the case of binomial SSD (see main text) as a function of the model parameters. The theoretical result (\ref{A.2.e}) is shown against numerical Monte Carlo simulation results using the Gillespie algorithm~\cite{Gillespie} and very good agreement is observed.

\vspace{4mm}

\begin{center}
	{\textit{3. bifurcation limit}}
\end{center}

\vspace{2mm}

The general result for the MTE [Eq.~(\ref{A.2.e})] can be drastically simplified close to bifurcation. Indeed, close to the bifurcation
the attracting fixed point $q_s=1-1/B$ is assumed to be close to $q=0$ such that $q_s\equiv\delta\ll 1$. As a result, we can denote $B\simeq 1+\delta$ such that $B-1\ll 1$. Using the equation $q_a(p_f)=0$, or $e^{p_f}f(p_f)=1/B$, assuming a-priori that $p_f={\cal O}(\delta)$ and expanding in powers of $\delta\ll 1$, we arrive at $p_f=\mathcal{S}'(0)\simeq-\delta/[1+f'(0)]$, thus justifying a-posteriori our assumption that $p_f={\cal O}(\delta)$. Performing a similar approximation to the rest of the terms in Eq.~(\ref{A.2.e}) we finally arrive at the MTE close to bifurcation
\begin{equation}
	\label{A.3.a}
	\tau=\frac{\sqrt{\pi(1+\gamma)}}{\sqrt{N}(B-1)^2}e^{\frac{N(B-1)^2}{1+\gamma}},
\end{equation}
where $\gamma$ is defined in Eq. (\ref{A.1.d}). The validity of the WKB approximation requires that $\mathcal{S}(0)\gg1$ and thus $N(B-1)^2\gg 1$ [$\gamma$ is assumed to be ${\cal O}(1)$]. On the other hand, in Eq.~(\ref{A.3.a}) we have accounted for the pre-exponential correction while neglecting terms on the order of $N(B-1)^3$ in the exponent. As a result, Eq.~(\ref{A.3.a}) is valid as long as $N^{-1/2}\ll B-1\ll N^{-1/3}$.

\vspace{6mm}


\begin{center}
	{\textbf{B. Birth-death-competition model -- mean time to extinction calculation}}
\end{center}

\vspace{4mm}

In this section, we present the derivation of the leading-order MTE for the birth-death-competition model presented by Eq.~(7) in the main text. The corresponding rate equation for this model reads
\begin{equation}
	\label{sb}
	\dot{\bar{n}}=\bar{n}\left(B-1-\frac{\bar{n}^{M-1}}{N^{M-1}}\right).
\end{equation}
Equation (\ref{sb}) yields two fixed points: a repelling extinction point, $n=0$ and an attracting fixed point $n_s=N(B-1)^{1/(M-1)}$. The dynamics of the model resembles the dynamics of the bursty reproduction model, where an initial population first relaxes to a long-lived metastable state, $n_s$, and then, after a long waiting time undergoes extinction via a large fluctuation.

Repeating the calculations done in Sec. A. 1 for this model, one obtains a stationary Hamilton-Jacobi
equation $\mathcal{H}(q,p) = 0$, with the Hamiltonian
\begin{eqnarray}
	\label{sc}
&&	\mathcal{H}(q,p) = q(e^p-1)\mathcal{H}_a(q,p), \nonumber\\
&&  \mathcal{H}_a(q,p)=B-e^{-p}+q^{M-1}\phi (p).
\end{eqnarray}
Here, we have defined $\phi(p)=[\sum_{k=0}^Me^{-kp}D(k)-1]/[\langle k\rangle(e^p-1)]$. The fluctuational momentum, $p_f$, satisfies the relation $\mathcal{H}_a(0,p_f)=0$ which yields $p_f=-\ln B$.

\begin{figure}[h!]
\centering
\includegraphics[trim = .95in 5.4in 1.5in 2.65in,clip,width=3.2in]{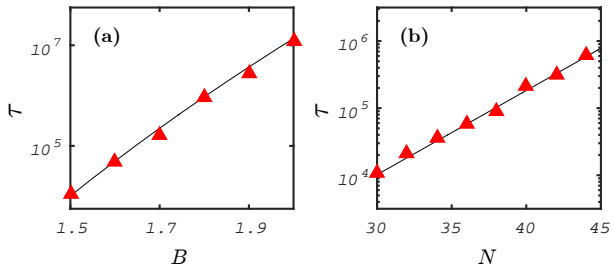}
\caption{(Color online) MTE for the birth-death-competition model, using a binomial SSD (defined in the main text), as a function of the model parameters. Lines correspond to the theoretical MTE, which is obtained by numerically evaluating the optimal path using Hamiltonian~(\ref{sc}). Markers correspond to numerical results using Monte Carlo simulations. Panel (a): $N=30$. Panel (b): $B=1.5$. Both panels: $K=3$ and $m=M=7$.}
\label{fig-sc}
\end{figure}

In order to obtain the MTE, we need to solve the equation $\mathcal{H}_a(q,p)=0$ (\ref{sc}) for either $q$ or $p$. In the general case it can only be done numerically. Analytical progress is only possible when one considers the bifurcation limit, where a solution can be found for an arbitrary distribution $D(k)$. Close to the bifurcation we denote $B-1\equiv\delta\ll1$, which yields $q_s=\delta^{1/(M-1)}$ and $p_f=-\delta+\mathcal{O}(\delta^2)$. Defining rescaled coordinates $\tilde{q}=q/\delta^{1/(M-1)}$ and $\tilde{p}=p/\delta$, and Taylor-expanding $\phi(p)$ in powers of $\delta$ up to subleading order, we find $\phi(p)\simeq\phi(0)+\phi'(0)p=-1+p(\gamma+1)/2$. Plugging this into $\mathcal{H}_a(q,p)$ [Eq.~(\ref{sc})] and using the rescaled coordinates we arrive at
\begin{eqnarray}
	\label{sd}
&&	\mathcal{H}_a^b(\tilde{q},\tilde{p})=\delta\left\{1+\tilde{p}-\tilde{q}^{M-1}\right.\nonumber\\
&&\left.-\delta\left[\tilde{p}-(\gamma+1)\tilde{q}^{M-1}\right]\frac{\tilde{p}}{2}\right\},
\end{eqnarray}
where the subscript $b$ means that we are close to bifurcation. Solving $\mathcal{H}_a^b(\tilde{q},\tilde{p})=0$ for $\tilde{p}$, perturbatively in $\delta$, we obtain the activation trajectory
\begin{equation}
	\label{se}
	\tilde{p}_a(\tilde{q})\simeq\left(\tilde{q}^{M-1}-1\right)\left[1-\frac{\delta}{2}\left(1+\gamma\tilde{q}^{M-1}\right)\right].
\end{equation}
As a result, the action function [Eq. (\ref{A.1.c})] in rescaled coordinates reads $\mathcal{S}(\tilde{q})=\delta^{\frac{1}{1-1/M}}\int_1^{\tilde{q}}\tilde{p}_a(\tilde{q}')d\tilde{q}'$, and thus, the MTE in the leading order is given by
$\tau\simeq e^{N\Delta\mathcal{S}}$, where
\begin{eqnarray}
	\label{sf}
&&\Delta\mathcal{S} = \left(1-1/M\right)(B-1)^\frac{1}{1-1/M}\nonumber\\
&&\hspace{8.10mm}\times\left[1-\frac{(B-1)}{2}\left(\frac{\gamma}{2M-1}+1 \right)\right].
\end{eqnarray}
This result holds as long as $N^{-1+1/M}\ll B-1\ll 1$, namely, $B$ cannot be too close to $1$. Importantly, this result indicates that at least close to bifurcation, the effect of reaction step-size noise in the bursty birth-death-competition model is introduced via a sub-leading term and thus, it is weaker than in the bursty reproduction model.

Figure \ref{fig-sc} presents the MTE for the case of binomial SSD (defined in the main text) as a function of the model parameters. One can see very good agreement between the theoretical and numerical results, where the latter are obtained using a Monte Carlo simulation~\cite{Gillespie}.

\begin{figure}[h!]
\centering
\includegraphics[trim = .95in 5.4in 1.5in 2.5in,clip,width=3.2in]{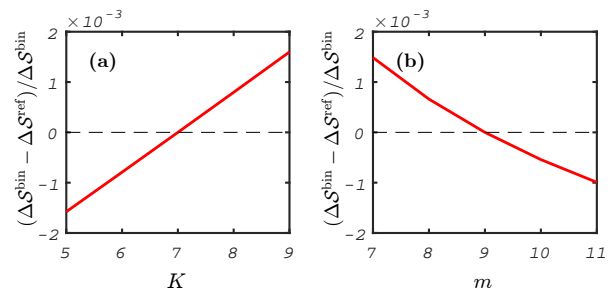}
\caption{(Color online) Relative difference between the accumulated actions associated with the binomial and reference SSDs as a function of their parameters, for the birth-death-competition model. Lines correspond to the theoretical result (\ref{sf}). Panel (a): $m=9$ (for which $K_\mathrm{c}=7$). Panel (b): $K=7$ (for which $m_\mathrm{c}=9$). Both panels: $B=1.15$ and $M=11$.}
\label{fig-sd}
\end{figure}

Figure \ref{fig-sd} shows the relative difference between the accumulated actions associated with the binomial SSD~(\ref{sf}) and with the reference SSD as a function of their parameters. This demonstrates that step-size noise can both increase or decrease the population's stability. In addition, it can be shown that the parameters of the birth-death-competition model can be tuned to maximize the effect of stability enhancement of the population, in a similar manner as shown for the bursty-reproduction model (see main text).

\begin{thebibliography}{99}
\bibitem{Langevin} This is similar to what happens when simulating a stochastic realization of a Langevin equation with delta-correlated Gaussian noise. Here, at each time step the noise magnitude is drawn from a given Gaussian distribution.
\bibitem{RE} The rate of the bursty reproduction was chosen such that the corresponding deterministic rate equation coincides with that of the usual logistic model, see below.
\bibitem{nonbifur} When $B-1\gtrsim {\cal O}(1)$, this procedure can only be done numerically by comparing the MTEs of the bursty and reference models. Yet, we have numerically demonstrated that the criterion for stability enhancement, $\gamma<\gamma^\mathrm{ref}$, valid close to bifurcation, also holds well into the non-bifurcation regime.
\bibitem{largem} Note that while the assumption $m\gg 1$ is reasonable, $m_{opt}$ can also be found for $m={\cal O}(1)$.
\bibitem{betabin} In the beta-binomial distribution, the success probability $p$ is not fixed, but is rather chosen from a beta distribution. This distribution, having three parameters, allows changing the skewness independently of the mean and variance, and thereby enables the demonstration of the stability-enhancement effect also for higher values of $R$.
\bibitem{Footnote1} Note, that here we use a different notation for the coordinate and momentum than in Ref. \cite{OURS-JSTAT}. Here $q$ and $p$ correspond to real space, and $\mathfrak{q}$ and $\mathfrak{p}$ correspond to momentum space, whereas in Ref.~\cite{OURS-JSTAT} $Q$ and $P$ correspond to real space, and $q$ and $p$ correspond to momentum space.
\bibitem{canonical} Here we have employed a canonical transformation of coordinates from the momentum space to real space, $\mathfrak{q}=qe^{-p}$ and $\mathfrak{p}=e^p$~\cite{Elgart,MS}.
\end{thebibliography}
\end{document}